\documentstyle[12pt,aaspp4,psfig]{article}
\input epsf
\input{amssym.def}      %
\input{amssym.tex}      
\def\kms{{\rm km/s}}

\def\cm2{{\rm cm ^{-2}}}
\def\mum{\mu {\rm m}}

\def\@journalname{Astrophysical Journal}
\def\accepted#1{\gdef\@accptdate{#1}}
\accepted{\relax}
\def\journalid#1#2{\gdef\@jourvol{#1}\gdef\@jourdate{#2}}
\def\articleid#1#2{\gdef\@startpage{#1}\gdef\@finishpage{#2}}
\begin{document}

\title{Detection of the 2175 \AA\ dust feature in Mg~II
absorption systems} 
\author {Sangeeta Malhotra}
\affil{IPAC, MS 100-22, California Institute of Technology,
Pasadena, CA~91125}

\begin{abstract}
The broad absorption bump at 2175 \AA\ due to dust, which is
ubiquitous in the Galaxy and is seen in the Magellanic clouds, is also
seen in a composite spectrum of Mg~II absorbers. The composite
absorber spectrum is obtained by taking the geometric mean of 92
quasar spectra after aligning them in the restframe of 96
absorbers. By aligning the spectra according to absorber redshifts we
reinforce the spectral features of the absorbers, and smooth over
possible bumps and wiggles in the emission spectra as well as small
features in the flat fielding of the spectra. The width of the
observed absorption feature is 200-300 \AA\ (FWHM), or 0.4-0.6
$(\mum)^{-1}$ and the central wavelength is 2240\AA{}.  These are
somewhat different from the central wavelength of 2176 \AA\ and
FWHM=0.8-1.25 $(\mum)^{-1}$ found in the Galaxy. Simulations show that
this discrepancy between the properties of the 2175 \AA\ feature in
Mg~II absorbers and Galactic ISM can be mostly explained by the
different methods used to measure them.

\keywords{ISM: dust, galaxies: intergalactic medium, quasars:
absorption lines}
\end{abstract}

\section{Introduction}

The amount of dust at high redshift has implications both for the chemical
evolution of galaxies and for observational strategies for detecting galaxies
at high redshift.  Quasar absorption systems offer an opportunity to study
dust in these systems at a large range of redshifts and hence trace the
chemical evolution.  Quasar absorption systems also have the advantages of a
simple geometry, being thin-screen absorbers backlit by quasars. Scattering
or more complicated effects of mixed dust and stars do not therefore
complicate the interpretation.

There are a few lines of evidence that indicate the presence of dust in
quasar absorption systems.  Fall, Pei \& McMahon (1989) and Pei, Fall \&
Bechtold (1991) compare the spectral slope of quasars with and without damped
Lyman-$\alpha$ absorbers and conclude that quasars with damped Lyman-$\alpha$
absorbers are redder. From the analysis of 100 and $60 \mum$ emission, Tanner
et al. (1996) conclude that quasars with Mg~II absorbers are redder in FIR
emission than random quasars. Meyer \& Roth (1990), Pettini \& Bowen 1997,
Pettini et al (1994, 1997) compare the observed gas phase abundance ratio of
Cr (which depletes onto dust) and Zn (which does not) in quasar absorption
line systems and conclude that a substantial fraction of the Cr must be in
dust grains (cf Lu et al. (1996), Kulkarni, Fall \& Truran (1997) for
discussion on nucleosynthesis patterns of elements relevant to such a
comparison).

Direct detection of spectroscopic features of dust would be more persuasive.
Absorption features can be an easier way to study the ISM of high redshift
galaxies than emission features: their strength is independent of the
absorber's redshift, depending only on the column density (of dust or gas)
and on the flux of the background source. The absorption bump at 2175 \AA\
is the strongest spectral dust feature in the ultraviolet [UV] - optical
wavelength range. The ratio of extinction at $2175 \AA$ to $A_V$ varies from
$A(2175)/A_V = $1.5-3.5 (Cardelli, Clayton, \& Mathis 1989), with the
higher value prevailing for dust in the diffuse medium.  McKee and Petrosian
(1974) first pointed out that the absence of the 2175\,\AA\ feature would
constrain the abundance of Galactic type dust in the high redshift damped
Lyman-$\alpha$ systems. Since then there have been a few studies looking for
the 2175\,\AA\ dust feature (Jura 1977; Smith, Jura \& Margon 1979; Boiss\'e
\& Bergeron 1988, Lanzetta, Wolfe \& Turnshek 1989, Pei, Fall \& Bechtold
1991). These studies constrain the 2175\,\AA\ feature to less than 1/10 the
Galactic strength in a few damped Lyman-$\alpha$ systems. This feature is
also seen at rest-wavelength of the quasar in PHL~938 and TON~490 (Baldwin
1977, Drew 1978)

The challenge in detecting the 2175\,\AA\ feature is that it is very
broad, FWHM$\simeq 350 \times (1+z)\AA $, and hard to distinguish from
broad undulations in the quasar emission spectrum. Coadding a large
number of quasar spectra after aligning them according to the absorber
redshift randomizes spectral features due to quasars and small
variations in the flat-fields, while reinforcing the features in the
absorption systems. I apply this technique to 96 Mg~II absorption
systems from the survey by Steidel \& Sargent (1992). The average
redshift of the Mg~II absorbers in this sample is $z \simeq 1.2$.

The main practical advantage of using the Mg~II sample lies in the
selection by the Mg~II doublet at 2800 \AA{}, close in wavelength to
the 2175\,\AA\ absorption feature, so that a large fraction of the
spectra from the survey had useful spectral coverage. Also, the
Steidel \& Sargent (1992) sample is a large, uniform survey with
absorbers at a wide range of redshifts ($0.2 < z < 2.2$). Most
previous searches of the 2175\,\AA\ feature were carried out for
damped Lyman-$\alpha$ systems, which have to be at $z > 1.65$ to be
observed from the ground. For high redshifts ($z > 2$) the 2175\,\AA\
feature lies in the noisier red end of the spectra. Also, one might
expect high redshift absorbers in a young universe (at $z \simeq 2$)
to be less chemically evolved.

The presence or absence of the 2175 \AA\ absorption feature can also help
distinguish between the various extinction curves that have been observed in
the Galaxy and the Large and Small Magellanic Clouds ( LMC and SMC). The
Far-UV extinction correlates with the properties of the $2175$ feature; the
broader the bump, the faster is the rise of FUV extinction (Fitzpatrick \&
Massa 1988). The presence of the 2175 \AA\ feature correlates well with
shallower far-UV rise of the extinction curve, implying less reddening (or
difference in extinction) between Visible and far-UV bands.  This is relevant
for high redshift objects, which are particularly vulnerable to
reddening/extinction because ground based optical observations correspond to
rest-wavelength UV (Ostriker \& Heisler 1984). The extinction and reddening
in UV is not only high, but also highly variable, The extinction at 1250 \AA\
may vary by 1.75 magnitudes between the bumpless SMC curve and the Milky Way
curve.

In \S 2 I describe the analysis methods to extract the composite absorber
spectrum. In \S 3 Monte-Carlo simulations are used to estimate the
significance of the detection of 2175 \AA\ feature in the composite
spectrum. \S 4 includes a comparison of the feature seen in absorbers with
that seen in the Galaxy and a discussion of dust-to-gas ratios and gas column
densities implied by the measurement of the 2175\AA\ feature.

\section{Analysis}

The composite absorber spectrum is derived from 92 quasar spectra from the
Steidel \& Sargent (1992, hereafter SS92) sample having 96 Mg~II absorption
systems. The red edge of the Lyman alpha forest absorption is bluer than the
2175 \AA\ feature in all systems where the 2175 \AA\ feature was in the
observed wavelength range. Visual inspection showed that about 45 absorption
systems were ``clean'' ; i.e. restwavelength $ 2175 \pm 200 \AA$ was in the
spectral range and did not coincide with emission features of the quasar. The
visual inspection was used only to estimate the number of absorbers
contributing to the signal and not as a basis of rejecting or accepting
spectra to be coadded.

First, the emission lines of the quasars are excised by removing data
points within 6000 $\kms$ of the following lines: Ly-$\alpha$ at
1215.7 \AA\, C~IV at 1549 \AA\, C~III at 1908.7 \AA\ and Mg~II at
2799.8 \AA{}. Then the composite spectrum is derived by taking the
geometric mean of the spectra in the SS92 sample after aligning them
according to the redshift of the absorption systems. Because of such
alignment any residuals in the spectrum which are correlated with
quasars redshifts (e.g. Fe~II emission) are randomized if the emission
and absorption redshifts are not correlated. The intrinsic quasar
spectra are multiplied by $(1+Z_{qso})/(1+Z_{abs})$ to shift them to
the absorber restframe. The distribution of $(1+Z_{qso})/(1+Z_{abs})$
is fairly featureless and decreases monotonically from 1 to 2.

If $f_{\lambda}$ is the intrinsic quasar spectrum and $\tau_i(\lambda)$ is
the dust absorption in the $i$th Mg~II system, the geometric mean gives
directly the average opacity due to dust.
$$(\Pi f_{\lambda} e^{-\tau_i(\lambda)})^{1/n} = (\Pi f_{\lambda})^{1/n}
\times e^{-\Sigma\tau_i(\lambda)/n} $$ In practice, the geometric mean was
calculated by running a boxcar mean on log($f_\lambda$).  Figure 1 shows the
composite spectrum using box-car of width 5 \AA{}. Different boxcar filters
were used and the final result was relatively insensitive to the width of the
filter for widths $< 20$ \AA{}.

\begin{figure}[htb]
\epsfxsize=7in\epsfbox{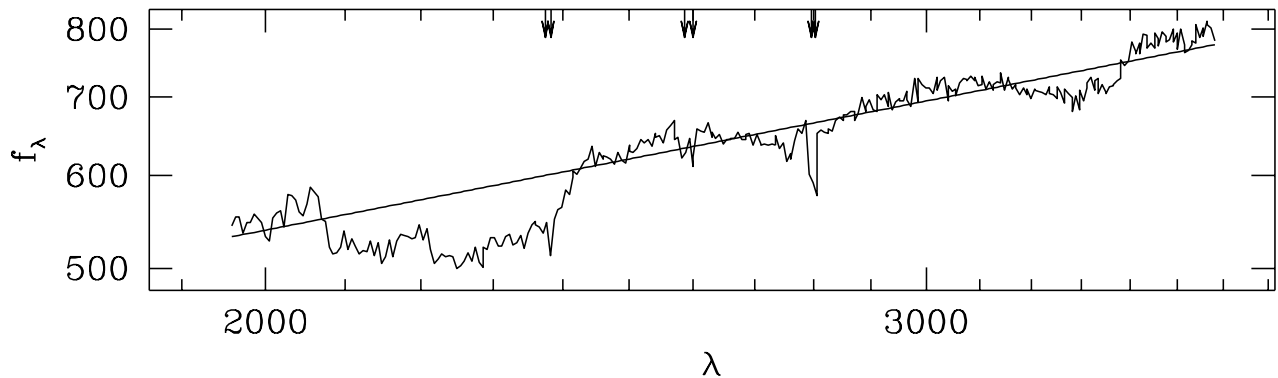}
\caption{The composite spectrum of quasar absorbers shown here is
obtained by removing quasar emission lines, aligning the spectra
according to the absorber redshift, and taking the geometric mean of
the spectra. The geometric mean is calculated using a boxcar mean of
width 5 \AA{} on Log$(f_{\lambda})$. The straight line is the
power-law fit to the spectrum at $\lambda > 2500 \AA$. The most
prominent feature in this spectrum is the absorption feature centered
at 2240 \AA{} which we identify to be the 2175 \AA\ dust feature seen
in the Galaxy . Other metal lines are also seen: the Mg II doublet at
2800 \AA\ and the Fe~II lines at 2600, 2586.65, 2382.77, and 2374.46
\AA{}.}
\end{figure}

\section{Error Analysis}

The absorption band seen at 2240 \AA\ is significant if only the
random noise is considered. In this section I will attempt to get a
more realistic estimate of the significance of this detection and to
test whether any systematic effects can spuriously produce this
feature. I test for the systematics in two different ways described
below.

\subsection{Randomizing the absorber redshifts}

To test how likely it was to get this feature spuriously, I repeated
the procedure of making the composite spectra, after randomly
shuffling absorber redshifts with respect to the spectra. The
amplitude of the composite spectra thus produced at 2240 \AA\ was
measured in 100 simulations. Only one simulation had a negative
deviation equal or exceeding the observed in the real spectrum. The
variance of the deviation at 2240 \AA\ in these random samples was
found to be 0.019. This implies that the dust feature in Figure 1 is
significant at $2.7\sigma$ level at peak (but there are about 40
points in the absorption feature at $> 2 \sigma$ level, so the total
significance of the detection is higher). Figure 2 shows the mean and
standard deviation of composite spectra in 100 such simulations.

\begin{figure}[htb]
\epsfxsize=7in\epsfbox{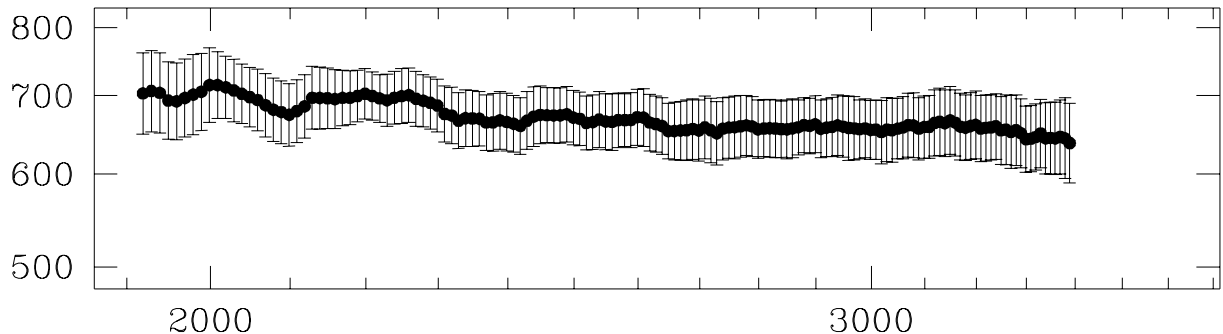}
\caption{ This is the resultant spectrum if we randomly mix the
absorber redshifts in the sample of SS92, and form a composite
spectrum in the same manner as in Figure 1. We see no features
comparable to the $2240 \AA$ dip in Figure 1. From the noise
estimated from 100 such composite spectra with randomized redshifts we
conclude that the peak deviation of the $2240 \AA$ dip in Figure 1 is
2.7$\sigma$ significant.}
\end{figure}

\subsection{Using a composite quasar emission spectrum}

It is possible that the absorption feature could be caused by small-amplitude
broad emission features in the quasar spectrum (e.g. FeII emission near 2400
\AA{}) combined according to the $Z_{qso}, Z_{abs}$ set in the SS92 sample
used.  To check that possibility I used a composite quasar spectrum without
absorption features to replace the real spectra in the SS92 sample.  This
spectrum was obtained by Zheng et al. (1997) by combining low redshift quasar
spectra observed with HST after removing the few foreground absorbers in
those spectra. Shifting and combining many copies of the Zheng et al. spectrum
using the same treatment applied to the SS92 sample yields a featureless
continuum (Figure 3).

\begin{figure}[htb]
\epsfxsize=7in\epsfbox{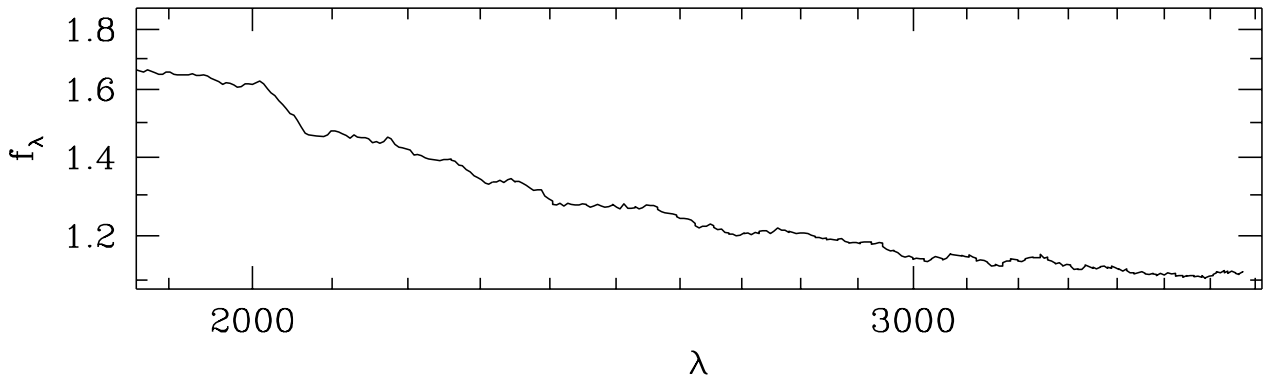}
\caption{The composite spectrum derived in the same manner as Figure
1, except the real spectra from SS92 have been replaced by a
low-redshift, high S/N quasar composite spectrum from Zheng et
al. 1997 which lacks absorption lines. The spectra are appropriately
redshifted to the absorber redshifts taken from SS92. We see no
feature comparable to the 2240 \AA\ absorption in Figure 1.}
\end{figure}

\section{Results and Discussion}

The most prominent feature in the composite spectrum obtained in Figure 1 is
the broad absorption bump with a central wavelength of 2240 \AA\ (or $ 4.46
\mum^{-1}$) and FWHM of 200-300 \AA\ (or $ 0.4-0.6 \mum^{-1}$). The peak
amplitude of the absorbing feature is 10\% (Figure 1).
Other narrow lines observed in the composite spectrum of the absorbers
are Mg doublet at at 2796 \& 2803 \AA\ and Fe~II lines at 2600,
2586.65, 2382.77 and 2374.46 \AA{}.

The interpretation of this absorption feature as the interstellar dust
feature seen at 2175 \AA\ in our Galaxy depends on whether it lies within the
range of observed extinction curves or theoretical models for dust.  Both are
somewhat uncertain. The carrier of this feature is most probably graphite or
amorphous carbon (Stecher \& Donn 1965, Gilra 1971, see review by Draine
(1989)). In the Milky Way this feature is observed to have a fairly constant
central wavelength $(\lambda_0=2176 \pm 9 \AA\ )$ (Fitzpatrick \& Massa
1986), while the width of the feature varies from $0.77 \mum^{-1}$ to $1.25
\mum^{-1}$.  Diffuse regions show narrower features, as do regions with
higher levels of radiation. This is consistent with the observed feature in
Mg~II absorbers being narrow, as Mg~II absorption comes from regions of
relatively low column density of neutral hydrogen. Although the central
wavelength of the feature associated with Mg~II systems is high for
interstellar dust, circumstellar dust in Hydrogen poor environments shows an
absorption bump at $\lambda_0=2400 \AA\ $ (e.g. Greenstein 1981, Hecht et
al. 1984, Drilling et al. 1997). The dust models have no difficulty
reproducing an absorption bump with $\lambda_0=2240$; such a feature can be
produced by including more large grains (Draine \& Malhotra 1993) or
amorphous carbon grains which produce an absorption bump at 2400 \AA{}.

To estimate the effect of the above methods of extraction on the
properties of the feature I perform Monte Carlo simulations in which a
dust absorption feature with known properties is artificially added to
spectra whose absorber redshifts are shuffled (as in \S 3.1).
Applying Galactic extinction curve with $A_V=0.3$, and with the 2175
\AA\ feature at central wavelength of $\lambda_0=2175 \AA$, FWHM=0.8
$\mum^{-1}$ and, I recover an absorption feature with absorption
amplitude 20\%, FWHM=0.5$\mum^{-1}$ and $\lambda_0=2230 \AA$ (Figure
4). This experiment shows that the procedure used to derive the
coadded absorption spectrum underestimates the width of the
feature. This may be because the feature is very wide compared to the
wavelength coverage (the FWHM of the feature $\sim 1/2$ the wavelength
coverage). The overestimate of the central wavelength of the feature
is presumably due to limited spectral coverage blueward of the
feature. The 2175 \AA\ feature in Galaxy, LMC and SMC is determined by
comparing the spectra of reddened and unreddened stars of the same
spectral type and does not suffer from these biases.

\begin{figure}[htb]
\epsfxsize=7in\epsfbox{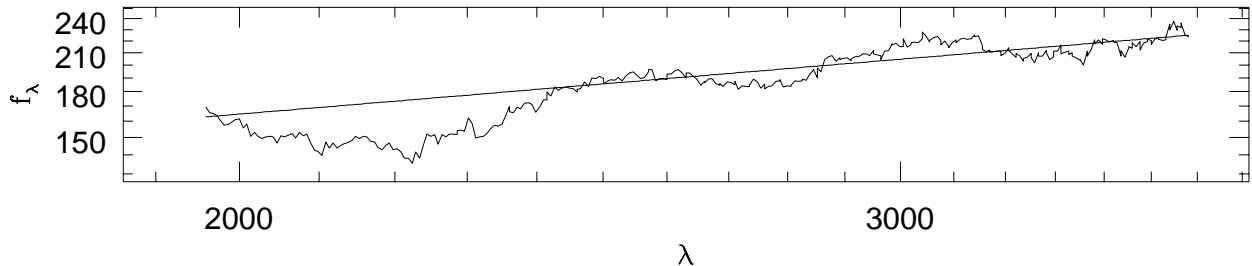}
\caption{This spectrum is derived by shuffling the quasars with
respect to the absorber redshifts so that the real absorption features
disappear, and then putting an artificial 2175 \AA\ feature into each
spectrum at the shuffled redshifts and coadding in the manner
described in section 2. This procedure enables a comparison of input
parameters of the absorption bump with the parameters derived by
subtracting a linear baseline and coadding several spectra. The FWHM
is underestimated, input FWHM$=0.8 \mum^{-1}$ yields
a FWHM$=0.5 \mum^{-1}$ in the final spectrum.}
\end{figure}

Because of the abundance variations and saturation of the Mg~II lines
at 2800 \AA\ , we cannot estimate the column density of hydrogen and
the gas-to-dust ratio in these systems. A peak amplitude of 10\% in
this feature of implies $A_V=0.15$ using the Galactic extinction law
(Cardelli, Clayton, \& Mathis 1989). Using the local gas-to-dust ratio
in the galaxy (Bohlin et al. 1978) - ${\rm N(H)/A_{v}=2 \times 10^{21}
cm^{2}}$, we derive an average $N(H)=3 \times 10^{20}$ for these
systems. Models show that Mg~II absorption systems are associated with
Lyman limit systems, i.e. $N(HI)> 10^{17.2}$ (SS92). The column
density distribution of Lyman limit systems is $N(HI)=B N^{-\beta} $,
with $\beta=1.2-1.7$( Steidel, Sargent \& Boksenberg 1988, Lanzetta
1991). Averaging over this distribution between column densities $
21.8 > log(N(HI))> 17.2$, yields an average column density $N(HI)
\simeq 10^{19}$. A lower limit to the average column density can also
be derived from the the fraction of damped Ly-$\alpha$ systems
($N(HI)> 2 \times 10^{20}$) among Mg~II systems.  This fraction is
estimated to be $0.13^{+0.29}_{-0.04}$ for $z\simeq2$ systems and $\le
0.14$ for $z\simeq 0.8$ systems (Rao, Turnshek \& Briggs 1995). The
damped systems alone would imply an average column density of $N(HI)
\ga 3 \times 10^{19}$. The column density of neutral gas should also
be corrected for ionized gas column density. The ratios of ionized to
neutral column density expected from photoionization model
$N(HII)/N(HI) \la 30$ for Lyman limit systems and and $N(HII) \simeq
N(HI)$ for $ N(HI)\simeq 10^{19}$ (e.g. Bergeron et al. 1994); this
may increase the average N(H) of the Mg~II systems. Molecular gas has
been found in a few absorption systems (Ge \& Bechtold 1997, Walker,
Bechtold, Black 1994). The strength of the 2175 \AA\ feature observed
is roughly consistent with Galactic type dust and dust-to-gas
ratio. Given the marginal detection of this feature and the
uncertainty in the column density of Hydrogen it is difficult to draw
any firm inference about the dust-to-gas ratio and the nature of the
extinction curve for these systems.

Deep imaging of Mg~II absorption systems has revealed them to be
normal star forming galaxies (Bergeron, Cristiani \& Shaver ~1992,
Steidel et al.~1994, Elston et al.~1991). It is not yet clear whether
the Mg~II systems arise from disks or halos of galaxies or from
material torn from galaxies in interactions (Churchill et al.~1997,
Bowen et al.~1996). If the damped Lyman-$\alpha$ systems are similar
to the Mg~II systems, a similar analysis should allow for an easier
detection of the 2175 \AA\ feature since the gas column densities are
higher. Possible detections of the 2175 \AA\ feature at quasar
redshifts have been reported (Baldwin 1977, Drew 1978), but attempts
to detect this feature in individual absorbers have failed (Jura 1977,
Smith, Jura \& Margon 1979; Boiss\'e \& Bergeron 1988, Lanzetta, Wolfe
\& Turnshek 1989, Pei, Fall \& Bechtold 1991), possibly because (a) it
has been looked for in high redshift damped Lyman-$\alpha$ systems,
which may be less evolved chemically and (b) It is difficult to
determine the intrinsic quasar continuum shape accurately enough to
detect an extinction feature of typical FWHM $\simeq 700 \AA$ in
observer rest frame. To be able to do that one may need to coadd many
quasar spectra as described in this paper.

\section{Summary and Conclusions}

We have evidence for the 2175 \AA\ dust feature in quasar absorption systems
identified by the Mg~II doublet at 2800 \AA{}. This feature implies the
presence of dust quite similar to Galactic dust. So far this feature had
been observed only in the local galaxies (Milky Way, LMC, SMC, M31, M101) where
individual stars can be resolved and their spectra compared to detect dust
absorption features in reddened vs unreddened stars of the same type. In
this study I have used the method of coadding many (96) spectra in the
absorber restframe to detect the wide (FWHM $\simeq 350 \AA$) dust
absorption feature. Most of the difference in properties of the bump
(e.g. FWHM) derived in quasar absorption systems and Galactic dust is shown
to be due to different methods used to derive these measurements.

\acknowledgements I am grateful to Drs Steidel and Sargent for making
their data available for this analysis.  I thank Ed Turner, James
Rhoads, Tomislav Kundi\'{c}, Buell Januzzi, and Roc Cutri for helpful
discussions and Drs Fall, Sargent and Boiss\'e for comments on an
earlier draft of the paper. I thank the referee for suggesting using
the geometric mean as a method to directly measure the optical depth
in the dust feature.

\end{document}